# Exponential Distance Relation and Near Resonances in the Trappist-1 Planetary System

Vladimir Pletser[1], Lorenzo Basano[2]

[1]Technology and Engineering Center for Space Utilization, Chinese Academy of Sciences, Beijing, China; Vladimir.Pletser@csu.ac.cn

[2] Department of Physics, Universita degli Studi di Genova, Genova, Italy; basano@fisica.unige.it

**Abstract**
We report in this paper a new exponential relation distance of planets in the newly discovered exoplanetary system of the Trappist-1 star, and we comment on near orbital mean motion resonances among the seven planets. We predict that possible smaller planets could be found inside the orbit of the innermost discovered Planet b.

*Keywords: Exponential distance relation; Titius-Bode law; Near mean motion resonances; Exoplanets; Trappist-1*

PACS: 95.10.Ce; 97.82.-j

## 1 Introduction

The number of exoplanets discovered recently has increased rapidly, mainly due to the Kepler mission and ground and orbital telescopes. One counts several thousands of exoplanets and thousands of multiple planetary systems. On the other hand, an abundant literature exists on the subject of the Titius-Bode law of planetary distances and their various generalizations and derivatives (see e.g. Nieto 1972). Based on the hypothesis of "holes" or "missing planets" (Basano and Hughes 1979), a revised form of the Titius-Bode law in an exponential form

$$a_n = \alpha \beta^n \qquad (1)$$

has been proposed thirty years ago (Pletser 1986), yielding the semi-major axis $a_n$ of the $n^{th}$ secondary in revolution around a central primary body, with $n$ integer and increasing radially outward from $n = 1$ for the innermost secondary, $\alpha$ and $\beta$ being real parameters, whose values for the Solar planetary system and the satellite systems of Jupiter, Saturn and Uranus are given in Table 1.

Table 1  Solar planetary and satellite systems

| Primary | $n_{max}$ | $\alpha$ | $\beta$ | LCC | Assumptions for $n_{max}$ |
|---|---|---|---|---|---|
| Sun | 9 | 0.214 AU | 1.711 | 0.9967 | 8 planets + Pluto[a] |
| Sun | 12 | 0.285 AU | 1.523 | 0.9987 | 8 planets + Pluto + 3 "holes"[b,c] |
| Jupiter | 14 | 0.658 $R_J$ | 1.567 | 0.9989 | 5 satellites + 2 sat. groups + 7 "holes"[c] |
| Jupiter | 13 | 1.017 $R_J$ | 1.571 | 0.9992 | 2ring groups+5sat.+2sat.groups+4"holes"[a] |
| Saturn | 25 | 0.841 $R_S$ | 1.251 | 0.9995 | 10 satellites + 15 "holes" [a,c] |
| Uranus | 8 | 0.871 $R_U$ | 1.522 | 0.9964 | pre-Voyager 2: 5 satellites + 3 "holes"[a,c] |
| Uranus | 8 | 1.143 $R_U$ | 1.456 | 0.9988 | post-Voyager 2: 2 ring groups+6 sat.[a,d] |
| Neptune | 15 | 1.124 $R_N$ | 1.422 | 0.9995 | post-Voyager 2: 6 satellites + 9 "holes"[a] |

$n_{max}$: maximum number of secondaries considered, including "holes"; units of $\alpha$ are AU (Astronomical Units) or equatorial radius of Jupiter ($R_J$), Saturn ($R_S$), Uranus ($R_U$), and Neptune ($R_N$); LCC: Linear Correlation Coefficient of linearized exponential regressions over semi-major axes in function of secondary attributed classification numbers; in Assumptions, groups of satellites or rings refer to a mean distance for that group used in the regression;
[a](Pletser 1990), [b](Basano and Hughes 1979), [c](Pletser 1986), [d](Pletser 1988)

The hypothesis of "holes" or "missing planets", originally applied to the Solar System (Basano and Hughes 1979) and later extended to satellite systems (Pletser 1986), was a very powerful tool when applied to the Uranian system before the Voyager 2 encounter. Assuming five Uranian main satellites (Miranda to Oberon) and three "holes", it allowed to predict distances at which a new ring and new satellites were discovered by Voyager 2 (Pletser 1986, 1988). A tentative reconstruction of a hypothetical satellite system of Neptune (Pletser 1989) allowed also to predict the radial distances of rings and small satellites discovered by Voyager 2 in 1989 (Pletser 1990).

Since the early nineties, more than several thousand exoplanets and exo-multiplanetary systems have been discovered. Attempts have been made to fit Titus-Bode-like relations to some of these exo-multiplanetary systems (see e.g. Lazio 2004; Christodoulou and Kazanas 2008; Poveda 2008; Qian 2011; Cuntz 2012; Altaie 2016; Aschwanden 2017). A study (Bovaird Lineweaver 2013) has been conducted in 2013 to see whether a generalized Titius-Bode law could be made applicable to those systems having at least four planets, to predict positions of yet undetected planets. This was challenged (Huang and Bakos 2014) as only some of the predicted planets have been discovered.

Recently, it was announced (Gillon et al. 2017) that the star Trappist-1 has a multiplanetary system composed of seven Earth-like planets arranged in a very regular configuration, on coplanar orbits with low eccentricities.

We report in this paper a new exponential relation among the semi-major axes of these seven planets. We comment also on the near orbital mean motion resonances among these planets and we predict that there might be one or several inner planets not yet detected.

## 2 Exponential distance relation

The observed semi-major axes $a_{obs}$ of the seven Trappist-1 planets (Gillon et al. 2017) are shown in Table 2, with two uncertainty ranges. The observed ratios $r_{obs}$ of successive semi-major axis $r_{obs.i} = a_{obs.i} / a_{obs.i-1}$ are shown in the next column, and it is seen that the system is very regular as all six ratios are close to the geometric mean of all ratios.

Table 2  Semi-major axes of Trappist-1 planetary system

| Trappist-1 | $a_{obs}$ ($10^{-3}$ AU) | $r_{obs}$ | n |
|---|---|---|---|
| b | 11.11 | -- | 1 |
| c | 15.22 | 1.370 | 2 |
| d | 21±6 | 1.380±0.394 | 3 |
| e | 28 | $1.333_{+0.534}^{-0.296}$ | 4 |
| f | 37 | 1.321 | 5 |
| g | 45 | 1.216 | 6 |
| h | $63_{-13}^{+27}$ | $1.400_{-0.289}^{+0.100}$ | 7 |
| Geometric mean | -- | $1.335_{-0.050}^{+0.015}$ | -- |

Therefore, there is no need to introduce any "hole" in the distribution and a linearized exponential regression on the observed semi-major axes (without uncertainty ranges) against classification numbers $n$ yields immediately a relation (1) with $\alpha = 8.7$ ($10^{-3}$ AU) and $\beta = 1.328$ which is also close to the geometrical mean of the ratios $r_{obs.i}$, and with a LCC = 0.9983.

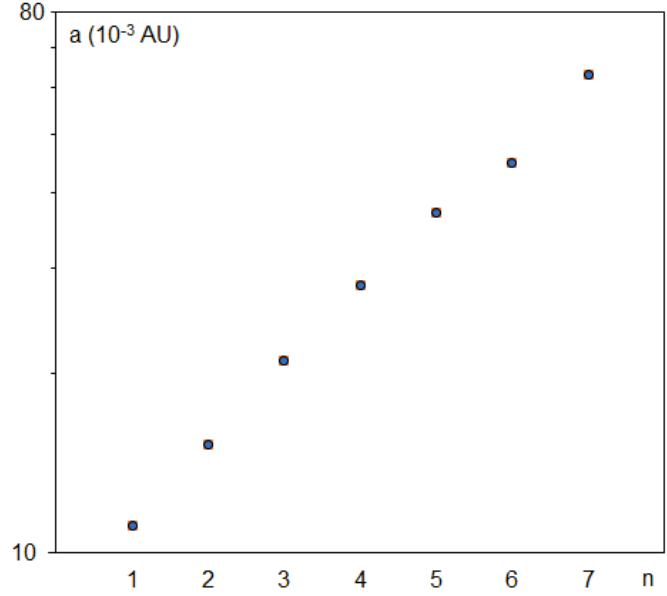

Figure 1: Semi-logarithmic plot of semi-major axes $a_{obs}$ of the Trappist-1 planetary system vs classification numbers $n$

## 3 Orbital Periods

Similarly, the observed periods $P_{obs}$ (Gillon et al. 2017) are shown in Table 3, with uncertainty ranges. The observed ratios $R_{obs}$ of successive periods $R_{obs.i} = P_{obs.i}/P_{obs.i-1}$ and the closest fractions of small integers are shown in the next two columns. For the last ratio of periods of Planets g and h, the uncertainty of the period of Planet h is so large that the ratio $n_h/n_g$ can vary between 9/8 and 17/6, with a middle value of 13/8.

Table 3   Periods of Trappist-1 planetary system

| Trappist-1 | $P_{obs}$ (days) | $R_{obs}$ | $n_i/n_{i-1}$ |
|---|---|---|---|
| b | $1.51087081 \pm 6.10^{-7}$ | -- | -- |
| c | $2.4218233 \pm 1.7\ 10^{-6}$ | $1.602932087 \pm 1.76174\ 10^{-6}$ | 8/5 |
| d | $4.049610 \pm 6.3\ 10^{-5}$ | $1.672132727 \pm 2.71872\ 10^{-5}$ | 5/3 |
| e | $6.099615 \pm 1.1\ 10^{-5}$ | $1.506222821 \pm 2.61491\ 10^{-5}$ | 3/2 |
| f | $9.206690 \pm 1.5\ 10^{-5}$ | $1.509388707 \pm 5.1812\ 10^{-6}$ | 3/2 |
| g | $12.35294 \pm 1.2\ 10^{-4}$ | $1.341735195 \pm 1.522\ 10^{-5}$ | 4/3 |
| h | $20_{-6}^{+15}$ | $1.619047773_{-0.485725341}^{+1.214313354}$ | $9/8 < 13/8 < 17/6$ |
| G. mean | -- | 1.538 | -- |

The periods follow also an exponential relation similar to (1), that reads

$$P_{calc.n} = 1.059*1.526^n \qquad (2)$$

with a LCC = 0.9977, and a value of $\beta$ = 1.526 close to the geometric mean 1.538.

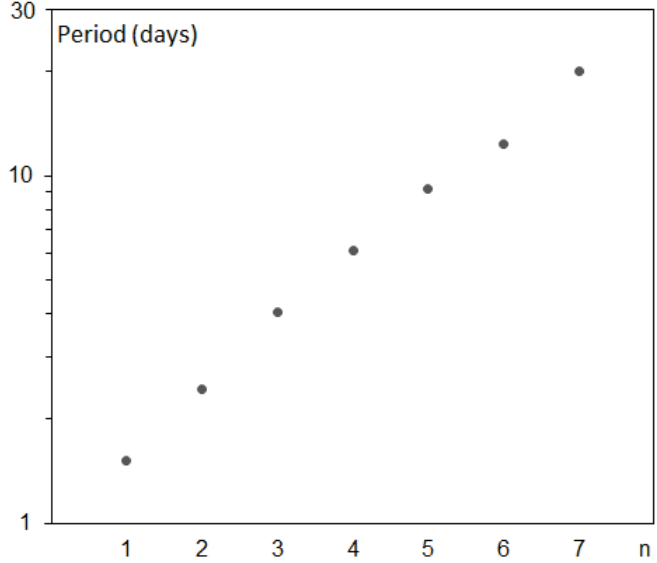

Figure 2: Semi-logarithmic plot of periods $P_{obs}$ of the Trappist-1 planetary system vs classification numbers $n$

Several conclusions can be drawn from this distribution.
First, the ratio of each orbital period to the previous one turns out to be close to the 3:2 mean motion resonance: the geometric mean for all planet's ratios is 1.538.
Second, although none of the planets are in an exact resonance, they are all very close to resonance, with, for Planets b to g, ratios of periods less than 1% of the ratios of small integers. Due to the large uncertainty range on the period of Planet h, it cannot be concluded whether it is close or not to a mean motion resonance with Planet g. However, if one consider the middle value of Planet h period, it is also close to a 13:8 resonance with Planet g. One would have to wait for the confirmation of Planet h's period value.
Third, except for the 4:3 near resonance between Planets f and g and considering the middle value of the integer ratio for the pair of Planets g and h, all two-body near resonances in Table 3 involve ratios of Fibonacci numbers, which is quite striking.
Fourth, the four innermost planets are close to a four-body near resonance
$$8:5:3:2$$
while Planets e and f and Planets f and g are close to resonances 3:2 and 4:3 respectively.
Fifth, if one includes also the Planet g next to the four-body relation, one has a five-body near resonance
$$8:5:3:2:1$$
One sees in Table 4 that the duration of respectively 8, 5, 3, 2, and 1 laps of Planets b, c, d, e, and g are in the range 12.087 to 12.353 days, i.e. with a difference of 2.2% or less.

Table 4  Five-body near mean motion resonances and duration in Trappist-1 planetary system

| Trappist-1 | $P_{obs}$ (days) | Number of laps | Duration (days) |
|---|---|---|---|
| b | 1.51087081 | 8 | 12.086966 |
| c | 2.4218233 | 5 | 12.109116 |
| d | 4.049610 | 3 | 12.148830 |
| e | 6.099615 | 2 | 12.199230 |
| g | 12.35294 | 1 | 12.352940 |

Sixth, if one includes also Planet f in this five-body relation, one has a weaker six-body near resonance,
$$24 : 15 : 9 : 6 : 4 : 3$$
but nevertheless, still very close as seen in Table 5, where the respective number of laps of the six planets have duration between 36.261 and 37.059 days, again with a relative difference of 2.2% or less.

Table 5   Six-body near mean motion resonances and duration in Trappist-1 planetary system

| Trappist-1 | $P_{obs}$ (days) | Number of laps | Duration (days) |
|---|---|---|---|
| b | 1.51087081 | 24 | 36.260899 |
| c | 2.4218233 | 15 | 36.327350 |
| d | 4.049610 | 9 | 36.446490 |
| e | 6.099615 | 6 | 36.597690 |
| f | 9.206690 | 4 | 36.826760 |
| g | 12.35294 | 3 | 37.058820 |

**4   Conclusions**

One can only be surprised by the regularity of this system. It shows further that Titius-Bode-like distance relations in exponential forms, or the equivalent for periods, are universal. One can now be persuaded that Titius-Bode-like relations have their physical roots in the dynamics of planetary systems and that they cannot be simply dismissed as numerology or be produced similarly by sequences of random numbers. There is an underlying physical explanation, either by a process of dynamical relaxation allowing planets to approach to mean motion resonances, or by responses of the initial protoplanetary disk submitted to radial perturbations. In this present case of the Trappist-1 system, the first explanation of coming close into resonance after dynamical relaxation should be considered.

Furthermore, the appearance of Fibonacci numbers is puzzling and the authors are wondering whether there might be more to it than simply a coincidence, possibly a configuration of minimum energy in the orbital mean motion resonances.

Finally, as the series of resonances 8:5:3:2:1 involves Fibonacci numbers, and as there is still room between planet b and the central star (its radius is $\approx 0.53 \cdot 10^{-3}$ AU), one hypothesizes that one or several smaller planets could be found between the star and the Planet b, and probably not yet detected as perhaps too small or too close to the star. The hypothetical inner closest neighbour to Planet b could have a period close to a 13:8 resonance with Planet b, such as to form a near resonance series 13:8:5:3:2:1. This would correspond to a period of approximately 0.9298 day and to a semi-major axis of approximately $\approx 8.37 \cdot 10^{-3}$ AU. Other inner planets could be found at, or close to, semi-major axes of approximately $\approx 6.3; 4.74; 3.57 \cdot 10^{-3}$ AU, ratios of which are equal or close to $\beta = 1.328$ or multiples thereof.

**Acknowledgement**

Vladimir Pletser is supported by the Chinese Academy of Sciences Visiting Professorship for Senior International Scientists (Grant No. 2016VMA042).